\documentclass[twocolumn,superscriptaddress,longbibliography]{revtex4-1}

\usepackage{graphicx}
\usepackage[a4paper,colorlinks=true,linkcolor=blue,citecolor=blue]{hyperref}
\usepackage[times]{mathastext}
\usepackage{float}
\usepackage{chngcntr}

\graphicspath{{figs/}{figsgaoerb/}}

\def\ket#1{\mathinner{|{#1}\rangle}}

\begin{document}

\title{Non-invasive magnetocardiography of living rat based on diamond quantum sensor}

\author{Ziyun Yu}
\affiliation{CAS Key Laboratory of Microscale Magnetic Resonance and School of Physical Sciences, University of Science and Technology of China, Hefei 230026, China}
\affiliation{CAS Center for Excellence in Quantum Information and Quantum Physics, University of Science and Technology of China, Hefei 230026, China}

\author{Yijin Xie}
\affiliation{Institute of Quantum Sensing and School of Physics, Zhejiang University, Hangzhou 310027, China}

\author{Guodong Jin}
\affiliation{CAS Key Laboratory of Microscale Magnetic Resonance and School of Physical Sciences, University of Science and Technology of China, Hefei 230026, China}
\affiliation{CAS Center for Excellence in Quantum Information and Quantum Physics, University of Science and Technology of China, Hefei 230026, China}

\author{Yunbin Zhu}
\affiliation{CAS Key Laboratory of Microscale Magnetic Resonance and School of Physical Sciences, University of Science and Technology of China, Hefei 230026, China}
\affiliation{CAS Center for Excellence in Quantum Information and Quantum Physics, University of Science and Technology of China, Hefei 230026, China}

\author{Qi Zhang}
\affiliation{CAS Key Laboratory of Microscale Magnetic Resonance and School of Physical Sciences, University of Science and Technology of China, Hefei 230026, China}
\affiliation{CAS Center for Excellence in Quantum Information and Quantum Physics, University of Science and Technology of China, Hefei 230026, China}

\author{Fazhan Shi}
\affiliation{CAS Key Laboratory of Microscale Magnetic Resonance and School of Physical Sciences, University of Science and Technology of China, Hefei 230026, China}
\affiliation{CAS Center for Excellence in Quantum Information and Quantum Physics, University of Science and Technology of China, Hefei 230026, China}
\affiliation{Hefei National Laboratory, University of Science and Technology of China, Hefei 230088, China}

\author{Fang-yan Wan}
\affiliation{CAS Key Laboratory of Brain Function and Disease, Hefei National Laboratory for Physical Sciences at the Microscale, Division of Life Sciences and Medicine, University of Science and Technology of China, Hefei 230026, China}

\author{Hongmei Luo}
\affiliation{CAS Key Laboratory of Brain Function and Disease, Hefei National Laboratory for Physical Sciences at the Microscale, Division of Life Sciences and Medicine, University of Science and Technology of China, Hefei 230026, China}

\author{Ai-hui Tang}
\affiliation{CAS Key Laboratory of Brain Function and Disease, Hefei National Laboratory for Physical Sciences at the Microscale, Division of Life Sciences and Medicine, University of Science and Technology of China, Hefei 230026, China}
\affiliation{Institute of Artificial Intelligence, Hefei Comprehensive National Science Center, Hefei 230088, China}

\author{Xing Rong}
\email{xrong@ustc.edu.cn}
\affiliation{CAS Key Laboratory of Microscale Magnetic Resonance and School of Physical Sciences, University of Science and Technology of China, Hefei 230026, China}
\affiliation{CAS Center for Excellence in Quantum Information and Quantum Physics, University of Science and Technology of China, Hefei 230026, China}
\affiliation{Hefei National Laboratory, University of Science and Technology of China, Hefei 230088, China}

\date{\today}



\begin{abstract}
Magnetocardiography (MCG) has emerged as a sensitive and precise method to diagnose cardiovascular diseases, providing more diagnostic information than traditional technology. 
However, the sensor limitations of conventional MCG systems, such as large size and cryogenic requirement, have hindered the widespread application and in-depth understanding of this technology. 
In this study, we present a high-sensitivity, room-temperature MCG system based on the negatively charged Nitrogen-Vacancy (NV) centers in diamond. 
The magnetic cardiac signal of a living rat, characterized by an approximately 20 pT amplitude in the R-wave, is  successfully captured through non-invasive measurement using this innovative solid-state spin sensor. 
To detect these extremely weak biomagnetic signals, we utilize sensitivity-enhancing techniques such as magnetic flux concentration. 
These approaches have enabled us to simultaneously achieve a magnetometry sensitivity of 9 $\text{pT}\cdot \text{Hz}^{-1/2}$ and a sensor scale of 5 $\text{mm}$. 
By extending the sensing scale of the NV centers from cellular and molecular level to macroscopic level of living creatures,  we have opened the future of solid-state quantum sensing technologies in clinical environments.

\end{abstract}

\maketitle
\newpage



\section*{Introduction}

Magnetocardiography(MCG) is a diagnostic technique that utilizes the cardiac magnetic signal to detect the heart activities\cite{ms1963detection}. 
In recent years, MCG has gained increasing attention for its higher sensitivity and accuracy in identifying cardiovascular diseases than conventional methods \cite{nomura1994noninvasive,koch2004recent,mori1988present,kandori2004identifying,van1999magnetocardiography,celermajer2012cardiovascular}. 
The cardiac magnetic signal propagates near-transparently through body tissue with low magnetic permeability, which allows the MCG method to detect more information than conventional techniques\cite{di2016future, nakaya1992magnetocardiography, alday2016comparison}. 
However, as the inherently weak strength of biomagnetic signals necessitate the high-sensitivity magnetometers in measurements\cite{kang2009measurement}, several critical challenges of these sensors need to be overcome for the widespread application of MCG technology. 
Among the conventional sensors, the most widely used superconducting quantum interference device(SQUID) is restricted by its expensive and complex cryogneic equipment\cite{strand2019low,koch2001magnetic, korber2016squids}. 
The thickness of the dewar in the SQUID magnetometer imposes constraints on achieving near-source room-temperature detection at millimeter scale, which is a critical limitation in biomagnetic applications\cite{wikswo1988high}. 
Recent advancement in atomic magnetometer is impressive, but its high performance is primarily  manifested in scalar magnetic field measurement\cite{li2018serf, wyllie2012optical, yang2021new, zheng2020vector, sheng2017microfabricated}, which causes a non-negligible vector information loss in magnetic cardiac measurement\cite{zheng2020vector}. 
As it stands currently, the challenging balance between sensor sensitivity, vector measurement capability and spatial resolution has hindered the widespread application of MCG technology\cite{fenici2005clinical, tavarozzi2002current}. 
This limitation from sensor also leads to the absence of high-resolution, full-vector cardiac magnetocardiography, 
which is the essential experimental result for gaining deeper understanding in this realm. 


The Nitrogen-Vacancy(NV) center in diamond is a promising solid-state spin system in quantum sensing\cite{degen2017quantum}, which is a competitively potential solution to overcome these challenges for MCG\cite{zhang2021toward, webb2020optimization}. 
This innovative quantum sensor has many advantages for magnetic field measurement, including high theoretical sensitivity,  \cite{taylor2008high,barry2020sensitivity, xie2021hybrid}, high spatial resolution\cite{hong2013nanoscale,glenn2017micrometer}, vector measurement capability\cite{zhang2018vector} and great integration potential in multi-channel array configuration\cite{zhang2021diamond,blakley2016fiber}. 
The dependency on cryogenic cooling in traditional SQUID-based MCG measurement is eliminated by the NV center's  high performance at room-temperature\cite{barry2016optical, zhang2021diamond}.
This advantage enables the diamond sensor to detect biomagnetic signals at micrometer-scale distances from source\cite{maletinsky2012robust}. 
Recently, a magnetometry sensitivity of 460 fT$\cdot$Hz$^{-1/2}$ has been achieved at the millimeter scale using NV centers\cite{barry2023sensitive}. 
Meanwhile, with the growing study of vector information's unique utility in diagnosing cardiovascular diseases\cite{zheng2020vector}, the NV centers demonstrate higher potential in MCG with their outstanding capability in vector field measurement\cite{schloss2018simultaneous}. 
These advantages position the NV center as one of the most competitive candidates to extend the frontier of MCG research and application. 

In this work, we performed the first instance of non-invasive MCG measurement of living animal using NV centers in diamond, which marks a meaningful milestone for this quantum system in biomagnetic sensing application. 
By applying sensitivity-enhancing techniques including magnetic flux concentration and dual-resonance spin magnetometry, we achieved a magnetometry sensitivity of 9 pT$\cdot$Hz$^{-1/2}$ in the 10 - 200 $\text{Hz}$ range. 
Based on the results of this animal experiment, the potential future enhancements and applications of diamond NV magnetometer in research and clinical setting are discussed in detail. 
\begin{figure*}
\includegraphics[scale=0.7]{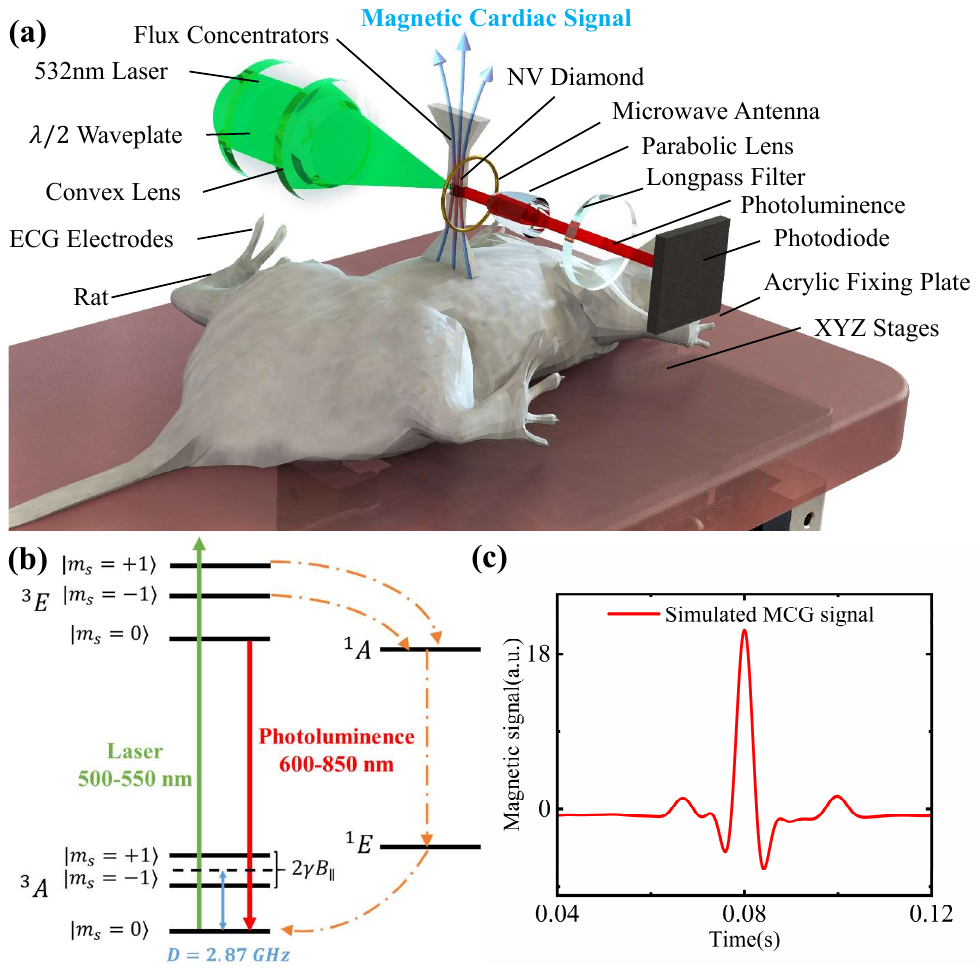}
\caption{\label{fig1}
Introduction of the Magnetocardiography(MCG) system based on NV magnetometer. 
(a) 3D schematic of the diamond-Based MCG system. 
The diamond-based MCG system detects the magnetic cardiac signal generated by experimental animal, which is significantly amplified by a pair of flux concentrators. 
An optical system was established for NV center excitation and red fluorescence collection, where the PL was sensed by a photodiode (PD) and readout by electronics devices. 
(b) Electronic structure and energy level diagram of the NV center in diamond. 
The zero-field splitting $D$ refers to the energy difference between the ground electronic spin state $\ket{m_s=0}$ and $\ket{m_s=\pm 1}$. When external field parallel to the NV symmetry axis $B_{\parallel}$ exist, there will be a Zeeman split between the $\ket{m_s=\pm 1}$ energy level which allows for the vector magnetic measurement. 
Green laser can be used to excite the NV center from ground state to the excited state, while the resultant red photoluminescence (PL) from spontaneous radiation serves as an optical readout for magnetic resonance.
(c) Simulated cardiac measurement experiment using the diamond-based MCG system. Here is the waveform of a  simulated cardiac measurement using this diamond-based MCG system.
A coil with two inversed ring radius $r_1= $5 $\text{mm}$ and $r_2=$ 6 $\text{mm}$ was fabricated with copper wire to approximate the rat heart at a speculated heart rate 12 Hz\cite{korecky1978normal}. 
The red line represents the simulated MCG waveform detected by this system, which evaluates the efficacy of this MCG system in detecting cardiac signal. 
}
\end{figure*}
\section*{Materials and Methods}
\begin{figure*}
\includegraphics[scale=0.65]{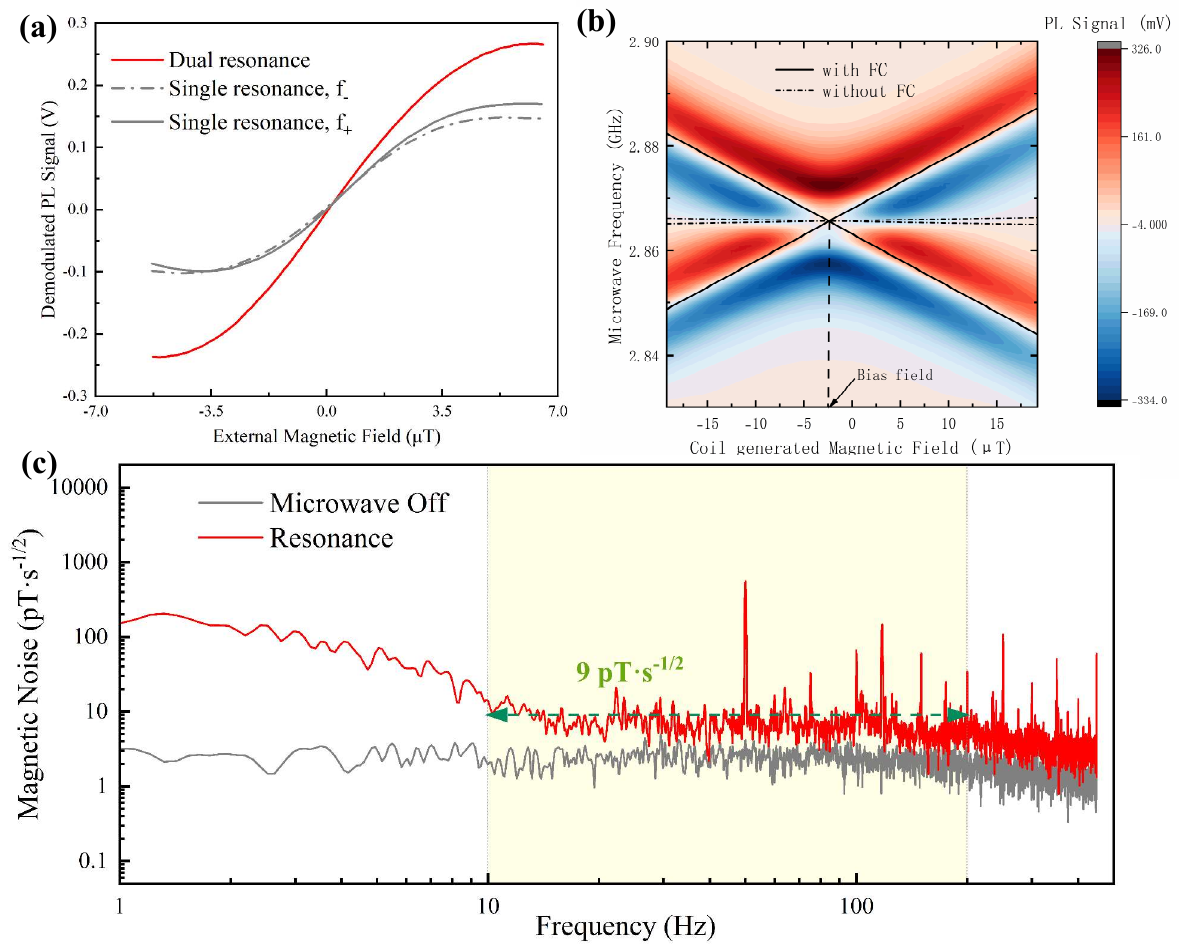}
\caption{\label{fig2}
Performance of the dual-resonance spin magnetometry based on NV center for the diamond-based MCG system.
(a) Dual-resonance magnetometry calibration.
The lines display the demodulated PL signal in response to varying external magnetic fields for both single and dual resonance methods. 
The red line illustrates the PL response in dual-resonance situation,  resonating simultaneously with both $f_{\pm}$ frequency with a linear response 72.6 $mV/\mu T$. 
The two grey lines represents the single resonance situations, each resonating with either $f_-$ or $f_+$ with linear responses 45.3 $mV/\mu T$ and 49.0 $mV/\mu T$.  
(b) 2D Optical detection magnetic resonance(ODMR) spectrum. The 2d ODMR spectrum of the NV centers in diamond demonstrates how the demodulated PL signal varies with an externally applied magnetic field along the flux concentrator and a changing microwave field. 
The solid line denotes the linear relationship between resonant frequency $f_{\pm}$ and $B_{ext}$ with magnetic flux concentration $f_{\pm}\approx D\pm \alpha G\gamma B_{ext}$, and the dotted dashed line represents the theoretical relationship without the flux concentrators $f_{\pm}\approx D\pm\alpha\gamma  B_{ext}$. 
Here $G$ is the enhancement factor of the magnetic flux concentrator, $\gamma$ is the gyromagnetic ratio of NV center, 
and $\alpha\sim 0.6$ is the angle factor from the misalignment between external magnetic field and the NV center's symmetry axis. 
As the slope of $\alpha G\gamma$ is fitted to be 1027$\pm$50 $\text{GHz/T}$ and $\gamma=$28 $\text{GHz/T}$, the amplification factor is estimated to be about $\sim 60$. 
(c) Diamond NV magnetometer sensitivity. 
This is the sensitivity spectrum of the NV magnetometer, where the red line represents the sensitivity of NV magnetometer, and the grey line indicates the background noise in non-resonant situation. 
The dotted green line represents the averaged sensitivity 9 pT$\cdot \text{Hz}^{-1/2}$ ranging from 10 Hz to 200 Hz, in accordance with the frequency range of cardiac signal. 
where the PL signal is sensitive to the environmental magnetic field. The experiment was carried out in the magnetic-shielded dru to ensure minimized environmental electromagnetic interference. 
}
\end{figure*}
\subsection*{NV center magnetometry}
NV center is a point defect in diamond consisting of a substitutional nitrogen atom and an adjacent vacancy. 
As shown in Figure \ref{fig1}b, the electronic spin of the NV center features a ground-state triplet $^3A$, where the lower energy state $\ket{m_s=0}$ is separated from the $\ket{m_s=\pm 1}$ states by a zero-field splitting D=2.87 $\text{GHz}$. 
When an external magnetic field $B_{\parallel}$ is applied along its symmetry axis, the ground-state $^3A$ splits due to the Zeeman effect\cite{rondin2014magnetometry}. 
NV center can be optically excited from the ground-state $^3A$ to the excited state $^3E$ using a green laser. After the excitation, its spin state can be optically readout through the red photoluminence(PL) intensity emitted during its spontaneous transition to the ground state\cite{acosta2013nitrogen}. 
This is attributed to a higher probability of a non-radiative transition via the intermediate singlet state $^1A$ and $^1E$ when the NV center is at $\ket{m_s=\pm 1}$ states than at $\ket{m_s=0}$ state.
The ground-state Hamiltonian of a single NV center can be written as followed, from which the state energy levels and transition  frequencies could be solved:
$$
H=D S_z^2+g\mu_B\vec B\cdot\vec S
$$
where $\vec S=(S_x,S_y,S_z)$ is the dimensionless electronic spin-1 operator with $\hat z$ parallel to the NV symmetry axis,  $\vec B$ is the external magnetic field while $B_{\parallel}$ is its component along the NV symmetry axis, $g$ is the electronic g-factor of the NV center, and $\mu_B$ is the Bohr magneton. The hyperfine coupling is ignored due to the broadened linewidth of the ODMR spectrum. 

When a continuous laser and a continuous microwave field resonant with the splitted spin energy level gaps are simultaneously applied to the NV center, the transitions between the $\ket{m_s=0}$ state and the $\ket{m_s=\pm 1}$ states are steadily induced. 
This process, known as microwave-induced spin resonance, influences the changes in PL intensity. 
Variation of the applied magnetic field alters the energy gap between the $\ket{m_s=0}$ state and the $\ket{m_s=\pm 1}$ states, thus changing the resonance situation for the microwave field. 
As a result, the PL intensity emitted by the NV center varies in response to the changes in the external magnetic field. 
By modulating the microwave field near the resonant frequency, the variation of PL intensity could be tuned into specific frequency and demodulated with lock-in technique\cite{barry2016optical}. 
When the microwave field applied is resonant with the electron transition, an observable change in the photoluminescence (PL) signal occurs. 
The degeneracy of $\ket{m_s=\pm1}$ is lifted when an external magnetic field is applied. 
This method ensures that the minute changes in external magnetic field could be accurately captured and analyzed, rather than being masked by low-frequency background noise. 

The dual-resonance spin magnetometry based on NV center is a technique to simultaneously manipulate the electron spin states $\ket{m_s=+1}$ and $\ket{m_s=-1}$ in external magnetic field. 
By modulating the two resonant frequencies at an identical modulation frequency $f_{mod}$ with $\pi$ phase shift, the overall PL signal could suppress common-mode noise and enhance sensitivity in single-resonance situation. 
The temperature-dependent zero-field splitting $D$ contributes to the resonant frequencies $f_{\pm}$ in common-mode behavior, while the magnetic field-dependent frequency shifting approximately $\pm\gamma B_{\parallel}$ affects $f_{\pm}$  in contrasting manner\cite{shin2012room}. 
As a result, this method effectively mitigates the influence of temperature fluctuations and improves the readout contrast for magnetic field detection.

\subsection*{System setup}
Figure \ref{fig1}a demonstrates the setup of the diamond-based MCG system for animal experiment. 
A diamond sample doped with ensemble NV centers was employed as the magnetic sensor. 
A pair of magnetic flux concentrators made from high magnetic-permeability material were fabricated to enhance the magnetic cardiac signal detected by this [100] diamond clenched between them\cite{fescenko2020diamond,xie2021hybrid}. 
The clamping surface size of the flux concentrator is 1 mm $\times$ 2 mm, and it is positioned at a distance of 10 mm from the bottom surface with size 5 mm $\times$ 10 mm.
As observed in the Figure \ref{fig2}b, the 2.5 $\mu T$ bias field is amplified by a factor of $\sim 60$ within the diamond sensor. 
To manipulate the spin state of the NV centers, a double split-ring microwave antenna was utilized to transmit the resonant microwave signal\cite{bayat2014efficient}. 
For a higher spin polarization of the NV centers, a 532 $\text{nm}$ green laser with 1.5 $\text{W}$ output power was focused to a beam waist of 40 $\mu m$. A half-waveplate was used to adjust the polarization of the excitation laser. 
The PL signal emitted from the NV centers was guided through a parabolic waveguide for higher collection efficiency. 
To ensure the purity of the optical signal, a longpass filter was used to exclude unwanted light components, allowing only the red PL to pass through. 
Finally, this refined red PL was converted into photocurrent by a photodiode(PD) for readout, which was then amplified by a pre-amplifier and received by a lock-in amplifier for signal detection. 
This setup ensures the diamond sensor's compatibility with the experimental animal for MCG measurement, as well as the efficiency of spin polarization and the collection efficiency of spin-emitted fluorescence. 

Following the setup of the MCG system, a simulated cardiac measurement experiment was performed using a homebuilt coil designed to mimic the shape and size of a rat heart\cite{kim2010measurement}. 
The coil's voltage, serving as the simulated ECG signal, was captured by a data acquisition system (DAQ) . 
Meanwhile, the generated magnetic field of the coil, representing the simulated MCG signal, was measured  by the diamond NV magnetometer. 
The MCG waveform obtained from this simulation is displayed in Figure \ref{fig1}c.
The figure clearly showcases the characteristics of a typical cardiographic waveform cycle (U, P, Q, R, S, T), effectively validated the capability of the MCG system in real magnetic cardiac signal measurement. 

To evaluate the practical performance of this MCG system, the magnetic responses of single-resonance magnetometry and dual-resonance magnetometry are measured, where the PL signals exhibit linear responses to the external magnetic field in a  near-resonant condition. This is illustrated in \ref{fig2}a. 
Meanwhile, Figure \ref{fig2}b exhibits the signal of modulated PL in varying magnetic field and microwave frequency. 
The microwave signal transmitted into the radiation structure consisted of two different NV resonant frequencies, which were utilized to perform the dual-resonance magnetometry. 
These two microwave signals, after being amplified, were combined and directed into the diamond. 
The sinusoidal modulation signal of the microwave was provided to the lock-in amplifier, serving as the external oscillator for PL signal demodulation. 
This configuration ensures precise synchronization between the modulation of the microwave signals and the detection process. 
The system also consisted of some auxiliary devices to perform MCG measurement. 
A pair of saddle-like coils were wired in the shielding cylinder to provide bias field and calibration waveform. 
The rat was set on a three-axis motion stage for position adjustment, which allows for a near approach to the diamond sensor. 
Three ECG electrodes were attached to the right forelimb, left forelimb and left hindlimb of the rat.
This arrangement allows different skin potentials to be measured, and thus the rat's electrocardiogram could be recorded in this lead-II configuration\cite{tontodonati2011improved}. 
The ECG data was recorded by the DAQ as reference for the MCG measurement. 
The MCG experiment was performed within a magnetic shielding cylinder made from multi-layer permalloy to suppress the electromagnetic noise in lab environment. 
The magnetic cardiac waveform was accumulated to enhance the signal-to-noise ratio.

\section*{Results}
\begin{figure*}
\includegraphics[scale=0.55]{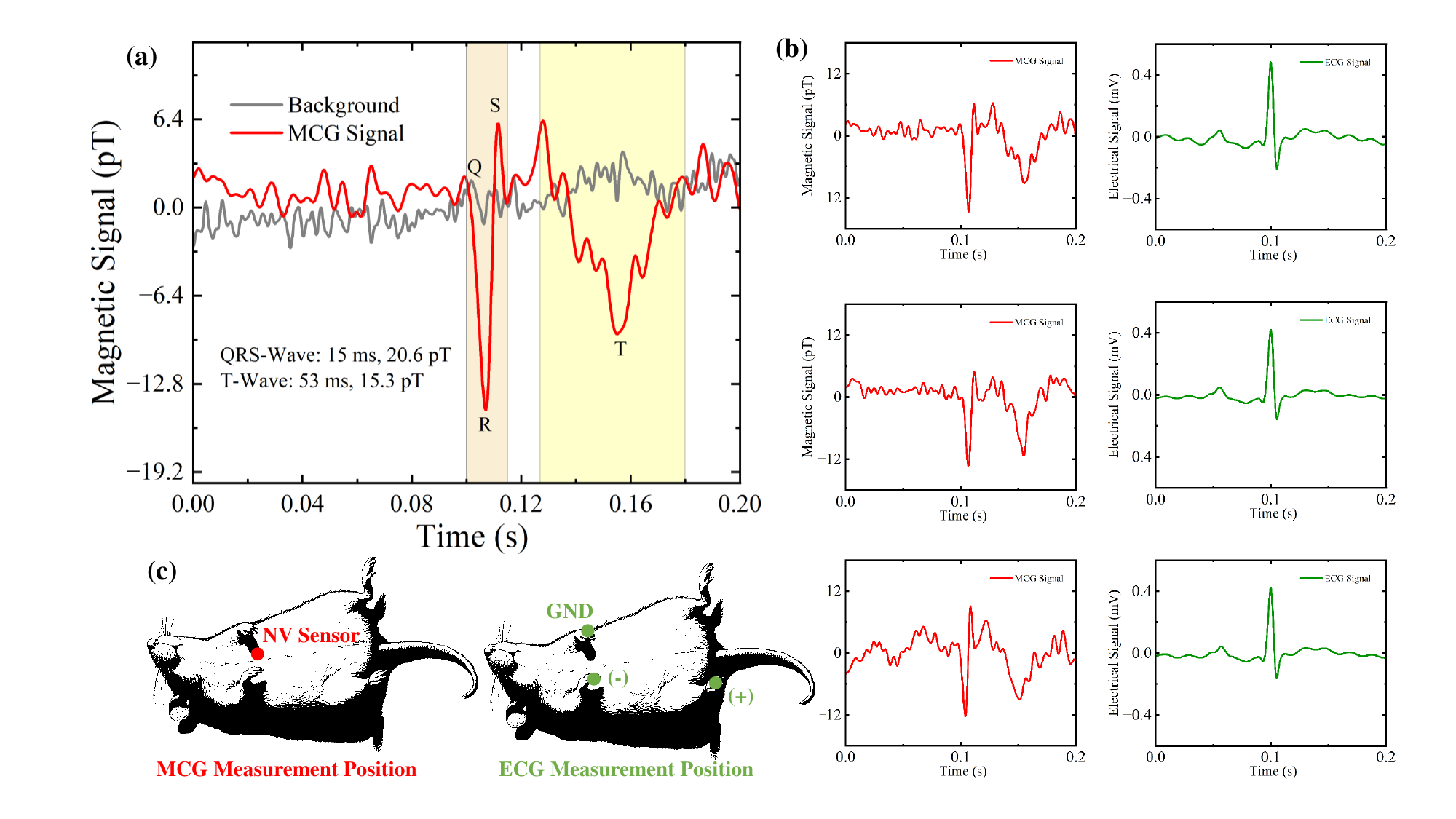}
\caption{\label{fig3}
Results of the non-invasive MCG experiment on rat. 
(a) Cardiac waveform acquired by diamond-based MCG system. The red line demonstrates the rat MCG waveform detected by the NV diamond sensor, while the grey line represents the background magnetic noise without the rat. 
The identified peak of typical Q, R, S, T wave are marked in the Figure. 
The orange area demonstrates the 15 $\text{ms}$ QRS-complex with amplitude $\sim$21 $\text{pT}$ in the rat MCG and the yellow area demonstrates the 53 $\text{ms}$ T-wave with $\sim$15 $\text{pT}$ due to the repolarization of the ventricles. 
(b) Comparison of rat MCG waveform and ECG waveform. The subplots with red lines represents the MCG waveforms acquired repeatedly in different experiments as in (a). 
And the green lines on the right represents the concurrently acquired rat ECG waveforms, which shows high conformity with the pertinent MCG waveforms. 
(c) Schematic of the MCG/ECG measurement. The MCG signal and ECG signal were captured simultaneously in the experiment. The left Figure demonstrates the setup of NV diamond sensor in the MCG experiment. 
The rat QRS-complex in ECG waveform is $\sim$0.62$\pm$0.7 $\text{mV}$. 
The diamond sensor clamped with the FCs was positioned right above the chest of the rat for 5 $mm$ from skin, 10 $cm$ on the spinal line from its nose to its root of tail. 
Meanwhile, as depicted in the right Figure, the ECG signal acquired by three electrode probes form the Lead II configuration of ECG measurement. 
}
\end{figure*}
\subsection*{MCG system calibration}

Before the animal experiment, we first assessed the background magnetic sensitivity of the magnetometer by conducting measurement without rat specimen. 
This result is presented in Figure \ref{fig2}b, where we plotted the density noise spectrum of the demodulated PL signal under both resonance and non-resonance situations by applying Fourier transformation of the data. 
In the frequency range related to the rat's cardiac signal spectrum, which spans from 10 $\text{Hz}$ to 200 $\text{Hz}$, the anticipated magnetic sensitivity is 9 pT$\cdot \text{Hz}^{-1/2}$.
As depicted in Figure \ref{fig3}a, after applying the accumulation method, the standard deviation of the background magnetic noise is about 1.6 $\text{pT}$. 
This result provides a benchmark for the following animal experiment.

\subsection*{Rat MCG data acquisition}
After the preparation of anesthetized rat, we positioned it near the diamond sensor and performed the non-invasive MCG measurements on this experimental animal. 
The ECG data were recorded in the lead-II configuration as shown in Figure \ref{fig3}c
For each test, the rat MCG data was acquired for about an hour and accumulated to enhance the signal-to-noise ratio. 
The ECG and MCG data were concurrently processed using a Butterworth bandpass filter, ranging from 10 $\text{Hz}$ to 200 $\text{Hz}$. Additionally, a 50 $\text{Hz}$ notch filter was specifically applied to these data.  
This dual filtering approach was performed to enhance the clarity and accuracy of the cardiac signal by excluding the noise out of interested frequency range and mitigating unshielded  electromagnetic interference. 
After filtering, the peaks of R-wave of ECG waveforms was utilized as a trigger in the signal accumulation process, where each peak of the R-wave was identified and marked by a threshold detection algorithm.
Subsequently, we first selected the time periods of equal length, which was chosen correspondingly to the cardiac cycle of rat (approximately 0.2 $\text{s}$), and centered these periods around the identified ECG peak timestamps. 
We then accumulated the rat MCG data within these selected time intervals to obtain the MCG signal.

A typical result of the accumulated MCG signal is displayed in Figure \ref{fig3}a. 
This figure clearly reveals magnetic cardiac activity signals, including the QRS-wave complex produced by ventricular depolarization and the T-wave resulting from ventricular repolarization.
Each of the stage is marked in the figure. 
The obtained MCG waveforms significantly surpass the background benchmark of 1.6 $\text{pT}$ measured in the absence of  experiment animal, and also demonstrates high consistency with the averaged ECG waveform which are indicated in the green lines of the subplots in Figure \ref{fig3}b. 
The experiment was repeated three times, yielding consistent results. 
The duration of the QRS-complex in the MCG waveform is approximately 17$\pm$2 $ms$, with an amplitude around 20.9$\pm$2.7 $\text{pT}$. 
The T-wave duration was approximately 41$\pm$10 $\text{ms}$ with amplitudes close to 15.4$\pm$0.7 $\text{pT}$.
The ST-segment, representing the interval isoelectric period between depolarization and repolarization, have a duration time length of about 17$\pm4$ $\text{ms}$. 
It is worth noting that the T-wave not obviously observable in the Lead-II ECG setup was successfully detected by the diamond MCG sensor, demonstrating the MCG method's capability to utilize more comprehensive information in cardiac activity monitoring. 
These results clearly showcased the capability to perform non-invasive MCG measurement with diamond magnetometer, providing a new solution to clinical MCG application with this solid-state quantum sensor.

\begin{figure}
\includegraphics[scale=0.4]{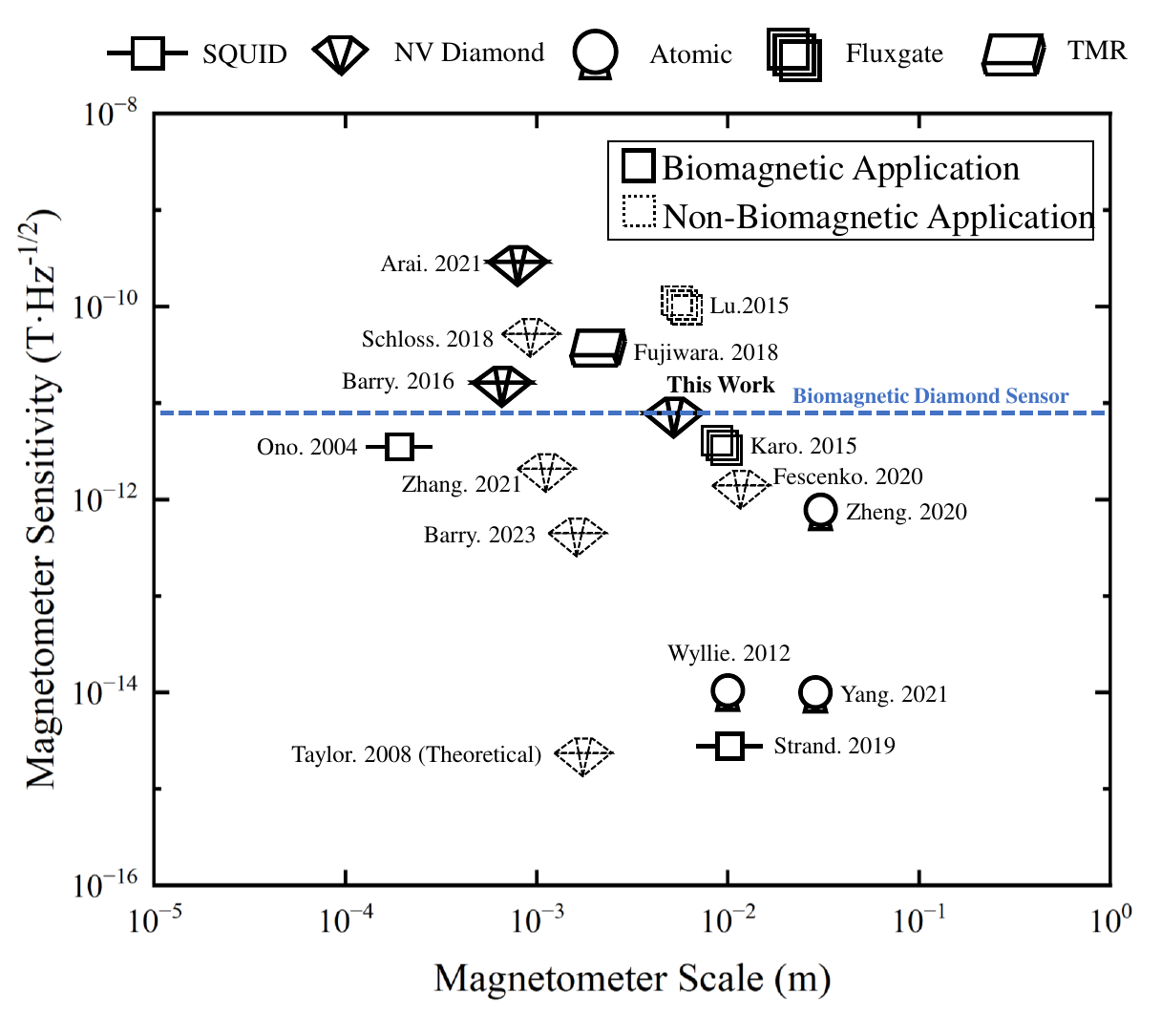}
\caption{\label{fig4}
A comparative demonstration of different magnetometers by scale and sensitivity, each type of sensor are denoted by different symbols.
The work related to biomagnetic measurements are depicted in solid lines, while the unrelated works are marked with dotted lines. 
Among all these sensors, the diamond NV center is notable for its high spatial resolution and high theoretical potential in full-vector magnetometry for millimeter-scale MCG\cite{taylor2008high,schloss2018simultaneous,barry2023sensitive,lu20153}. 
The blue dotted line represents the sensitivity of diamond NV magnetometers applied in biomagnetic measurements. 
One of the highest resolution MCG system was developed for mice cardiac measurement \cite{ono2004development}, but the minimum detection range limited by dewar thickness restricted the application of SQUID in sub-millimeter scale biomagnetic measurement\cite{strand2019low}. 
This work extends the application realm of NV diamond magnetometer in macroscopic living animal level.  
}
\end{figure}

This work commits the first instance of non-invasive MCG measurement using diamond NV magnetometer.
The improved sensor performance achieved through sensitivity-enhancing techniques has eliminated the need for destructive thoracotomy surgery, which was previously required to shorten the distance from the source\cite{arai2021millimetre}. 
This advancement marks the milestone in biological application of this solid-state quantum sensor, extending its scope from conventional role as a nano-scale spin sensor in microscopic imaging to a groundbreaking application in the practical realm of non-contact, non-invasive magnetic cardiac measurements for clinical purposes\cite{hong2013nanoscale,wang2021nanoscale,maze2008nanoscale}. 
Given that the human MCG signal is much stronger in signal intensity and larger in source scale than rat\cite{uchida2000current,kim2010measurement}, the successful rat experiment in this work provides substantial evidence supporting the feasibility of applying this room-temperature quantum sensor in clinical settings\cite{ms1963detection,kang2009measurement}. 
 
On the basis of this work, the performance of the ensemble NV center magnetometer could still be further optimized for higher measurement efficiency in clinical MCG application. 
The employment of cutting-edge technologies in high-quality ensemble NV diamond fabrication have achieved the magnetometry sensitivity of 460 fT$\cdot \text{Hz}^{-1/2}$, which indicates a potentially higher performance than we have achieved in this work\cite{barry2023sensitive}. 
For the magnetometry method we used, the estimated sensitivity could be depicted as followed\cite{xie2021hybrid}:
$$
\eta=\frac{\delta S}{G\alpha\gamma S_v\sqrt{2f_{ENBW}}}
$$
From which the $G$ is the amplification factor of the magnetic flux concentrator, $\delta S$ is the minimum detectable signal in non-resonant condition, $S_v$ is the maximum slope of the 1st-order derivative CW-ODMR spectrum, $\gamma$ is the gyromagnetic ratio of the electron spin, $f_{ENBW}$ is the equivalent noise bandwidth of the system, and $\alpha$ is the angle factor from the misalignment between external magnetic field and the NV center's symmetry axis. 
By optimizing the configuration of the diamond-flux concentrator and reducing electronic noise, we aim to further enhance the measurement sensitivity based on the results of this work. 
The simulation results are described in the Figure \ref{SM_fig2} in the Appendix D. 
Considering that $G$ increases with the shortening of the gap width between the flux concentrators within a certain range\cite{fescenko2020diamond}, a finite element simulation was conducted to establish the relationship between these two factors in current configuration. 
Additionally, considering the negative impact of this configuration change and reduced air gap on the efficiency of fluorescence collection, optical simulations were also performed for fluorescence collection efficiency based on a uniform excitation and point-source model.
To augment the signal strength, the simulations included a reflective coating on the clenching-face of the flux concentrator and an additional reflective mirror for fluorescence reflection\cite{yu2020enhanced}. 
Also, polishing the clamping surface of the flux concentrators could reduce the magnetic inhomogeneity in the diamond and narrow the ODMR spectrum's line shape, thereby increasing the magnetometry sensitivity by improving the CW-ODMR slope $S_v$\cite{xie2023towards}. 
In the preliminary simulations presented in the Appendix D, we observed promising total prospects for sensitivity enhancement in our system. 
Specifically, by reducing the gap width and optimizing the photoluminescence (PL) collection scheme, we anticipate a threefold increase in sensitivity. 
Additionally, a 3.5-fold enhancement in sensitivity is expected from the reduction of background noise. These improvements, as evidenced by our simulations, hold significant potential for advancing the performance of our system.
As a result, we may anticipate a sub-picotesla level magnetic sensitivity by applying optimization methods. 

This research implies a widespread practical application MCG based on diamond NV centers in clinical settings.
Also, it lays the groundwork for future explorations in biomagnetic measurement using quantum sensors. 
Traditional MCG maps are constrained by intrinsic sensor limitations, which are usually in the form of single-axis or scalar measurements\cite{yang2021new, brisinda2004clinical, gusev2017ultra}. 
These conventional methods also grappled with constraints in sensor size, resolution and minimum detectable distance, restricting the spatial intervals of MCG sampling positions at a level of a few millimeters\cite{ono2004development}, leaving the unknown scale lower than that to explore. 
However, the diamond crystal structure of the NV center allows a small-sized single diamond sensor to perform  precise full-vector magnetic measurements\cite{schloss2018simultaneous,zheng2020vector}. 
Together with this advantage, the NV center also stands out with its high-resolution capability, functioning as an sensitive magnetometer at the sub-millimeter scale, which is firstly attempted in this previous work as cited\cite{arai2021millimetre}. 
This capability could be utilized to generate high-resolution full-vector biomagnetic maps at sub-millimeter scale - a realm rarely attained previously. 
With complete magnetic information of cardiac activity accessible, the limitations of current MCG maps might be overcome, facilitating more specific applications such as precise localization and inversion\cite{schmidt1986multiple}, biomagnetic mechanism modeling\cite{comani2004independent,brisinda2004multichannel,comani2004concentric}, and the potential higher diagnostic performance based on the full-information MCG data\cite{Meyerfeldt1999ICB,nikitin1996magnito}. 

In conclusion, we successfully presented a MCG system based on ensemble NV centers in diamond. 
The first instance of non-invasive MCG measurement on living animal using NV centers was successfully performed using this diamond-based MCG system. 
Furthermore, a prospective analysis of diamond NV magnetometer's potential in the field of biomagnetic sensing was conducted to discuss the future development of this quantum sensor. 
This work marks a meaningful milestone for diamond-based MCG system in both clinical setting and biomagnetic research. 

\section*{Acknowledgments}
This work is supported by the National Key R\&D Program of China (Grants No. 2021YFB3202800, No. 2022YFF1400400 and No. 2018YFA0306600),
the Chinese Academy of Sciences (Grant No. GJJSTD20200001), 
Innovation Program for Quantum Science and Technology (Grants No. 2021ZD0302200, and No. 2021ZD0303204),
the Fundamental Research Funds for the Central Universities (Grant No. WK3540000011, 226-2023-00137, 226-2023-00138), 
and Anhui Initiative in Quantum Information Technologies (Grant No. AHY050000). 
X. R thank the Youth Innovation Promotion Association of Chinese Academy of Sciences for the support. 

\section*{Author contributions}
Z.Y., Y.X, and G.J. contributed equally to this work. 
Z.Y. completed the simulation calculations and experimental design, performed the experiments and data analysis with G.J..
X.R. proposed the idea of this work. 
Y.X and Y.Z completed the preliminary simulated experiments.
Q.Z., F.S., F.W., H.L. and A.T. provided assistance for the animal experiment. 
All authors contributed to the writing of the manuscript. 

\appendix
\counterwithin{figure}{section}
\section{Microwave and electronical acquisition system}

\begin{figure*}
	\includegraphics[scale=0.6]{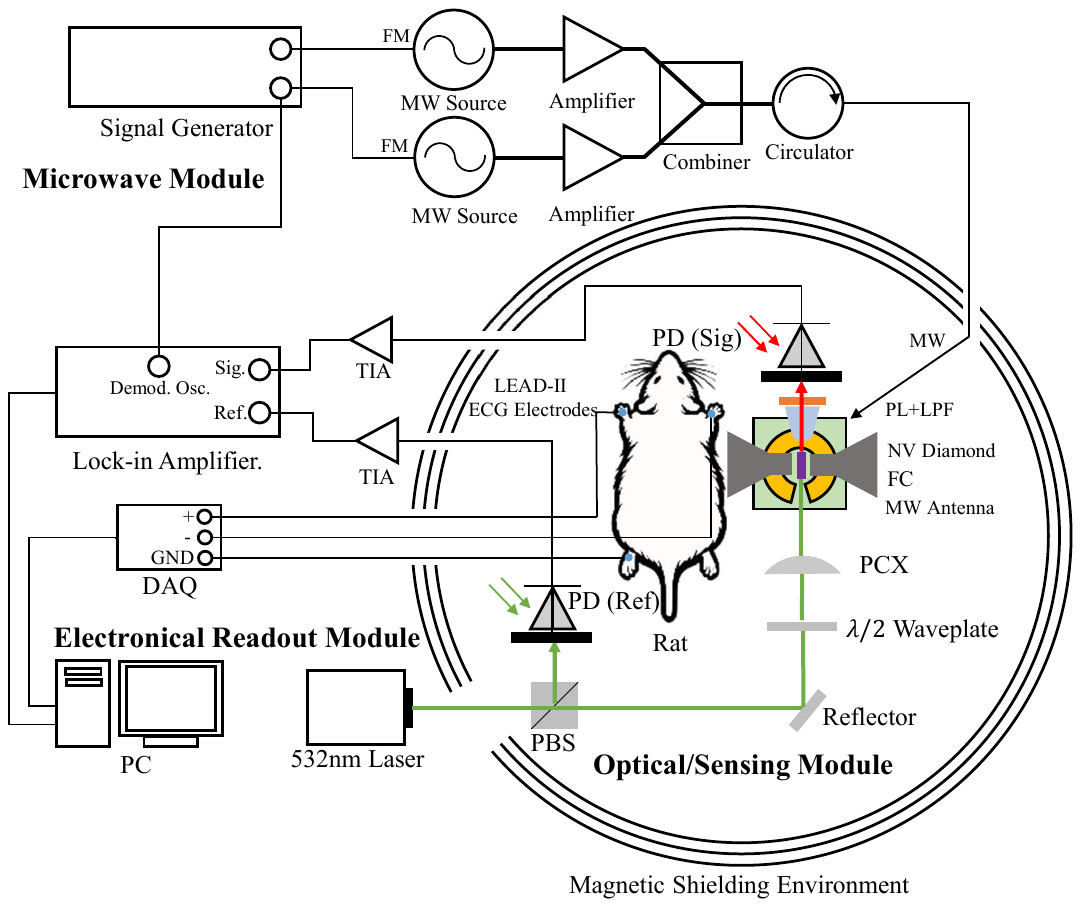}
	\caption{\label{SM_fig1}
		Setup diagram of the diamond-based MCG system.
		The system consisted of three primary modules. 
		The first was a microwave module that generates the microwave signal to manipulate the NV spins in diamond via dual-resonance method.
		The second was an optical/sensing module tasked with exciting the NV centers and collecting the PL emitted from them, together with auxiliary electrodes for ECG measurement.
		The third was and an electrical readout module to capture and synchronize both the MCG signal and ECG signal. 
		Two home-built trans-impedance amplifiers (TIA) were employed to extract and amplify the AC fluorescence photocurrent by a factor of 2000 $V/A$. 
	}
\end{figure*}

The microwave system consists of two FSW-0010 sources which generate microwave signal at two different NV resonant frequencies to perform dual-resonance magnetometry, as depicted in the Figure \ref{SM_fig1}.
A double split-ring structure was fabricated to apply the microwave field to the diamond\cite{bayat2014efficient}. 
The two microwave outputs are modulated by the same optimized frequency of 21637 Hz with a $\pi$ phase shift produced by a signal generator Rigol DG-812, so that the demodulated PL signal from $\ket{m_s=+1}$ and $\ket{m_s=-1}$ states could reserve the common mode magnetic signal and eliminate the anti-common mode temperature signal. 
This dual-resonance magnetometry method was deployed to simultaneously manipulate the electron spin resonant levels $\ket{m_s=+1}$ and $\ket{m_s=-1}$ in external magnetic field, effectively minimizing the impact of temperature shift and improving the readout contrast for magnetic field\cite{fescenko2020diamond}.

\begin{figure*}
	\includegraphics[scale=0.7]{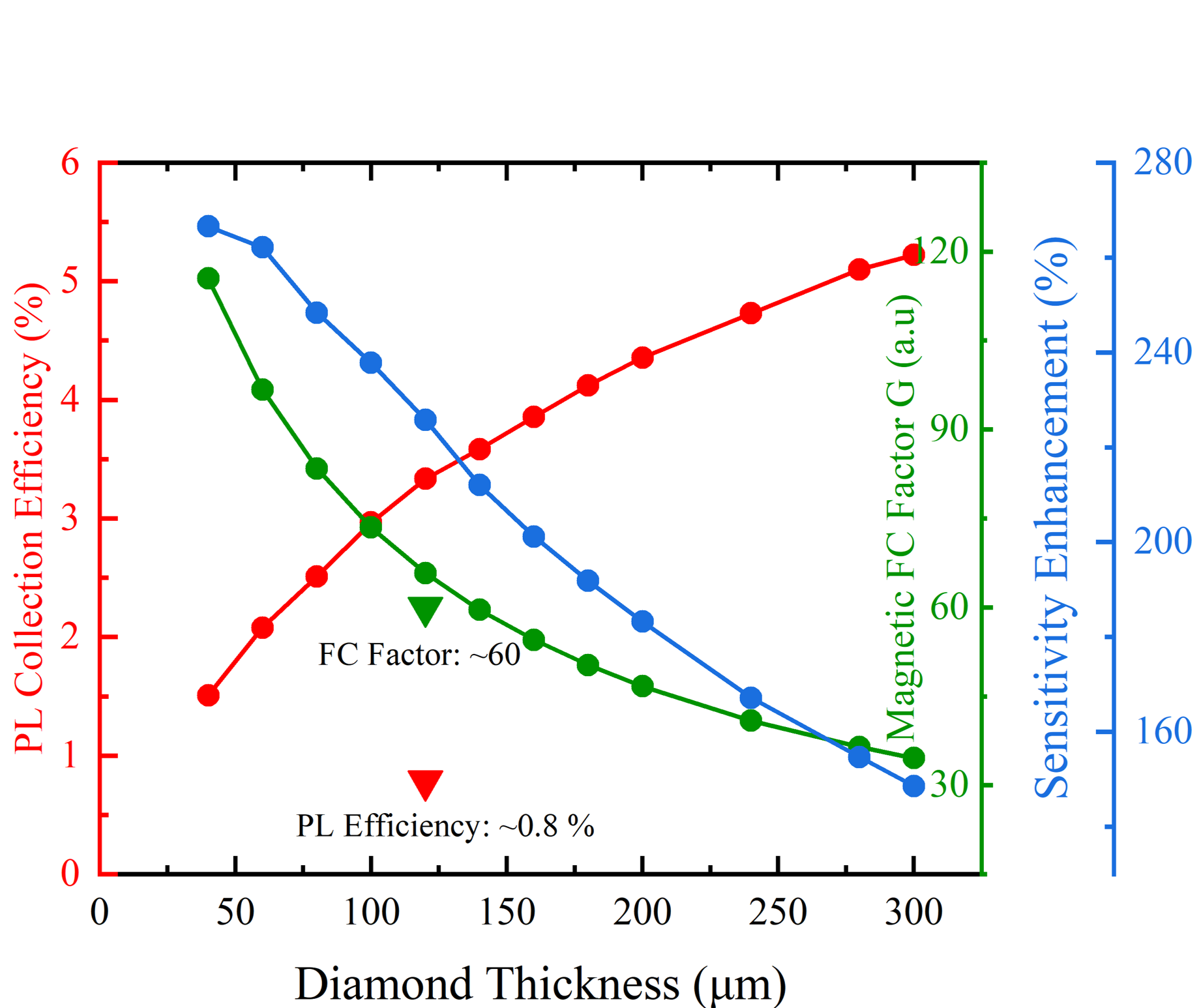}
	\caption{\label{SM_fig2}
		Simulation of the potential sensitivity enhancement. 
		This figure demonstrates the relationship between diamond thickness's relationship, PL collection efficiency and magnetic flux concentrator's amplification factor G. 
		the blue line graphically denotes the expected sensitivity enhancement from simulation, while the green and red lines represent G and the PL collection efficiency respectively. 
		Here the sensitivity enhancement is directly proportional to G and relates to the square root of the phonon number $\sqrt{N}$. 
		The current system parameters are indicated in the graph by the inverted triangles.
		To augment the signal strength, the simulations included a reflective coating on the clenching-face of the flux concentrator and an additional reflective mirror for fluorescence reflection\cite{xie2023towards,xie2021hybrid}. 
		The best expected enhancement about 3-fold is achieved at 40 $\mu m$ gap width, where the magnetic amplification factor $G\approx120$ and the fluorescence collection efficiency $\approx$1.5\%. 
		Combining the 280\% improvement from the system optimization and the 350\% potential improvement from the background noise reduction (2.6 pT/$\text{Hz}^{-1/2}$), 
		we anticipate an approximate tenfold enhancement in magnetometry sensitivity based on the system's current performance. 
	}
\end{figure*}\textbf{}

In the development of the electronic acquisition system, we employed a phase-locked loop readout technique based on microwave frequency modulation method, which could effectively avoid the 1/f electronic noise. 
The microwave-modulated PL signal and reference laser signal are separately converted into photocurrent with two silicon photodiodes(PD), and demodulated with a 2-channel lock-in amplifier HF2LI. The magnetic field-dependent PL signal acquired by the lock-in amplifier. 

\section{Optical System}
The optical system comprises of a laser excitation module and a fluorescence collection module, as depicted in the Figure \ref{SM_fig1}.
A Cobolt-Samba-1500-532nm laser with output power 1.5 W is used to generate the green light and excite the NV centers in diamond.
The laser is sampled by a polarized beam splitter (CCM1-PBS251/M), which allows the laser's reference photoelectric signal to be acquired by the lock-in amplifier to cancel the laser power fluctuation. 
The laser beam was adjusted by several reflectors and focused with a spherical lens with focal length f=20 $\text{mm}$ (LA1074) to achieve maximum laser power density in diamond. 
An NV diamond was fabricated as the sensor of this system, it is type-Ib, CVD-grown, [100] cut, measuring 1 mm x 1 mm x 0.12 mm, with an NV density of approximately 0.3 ppm.
The angle between the magnetic flux concentrator's orientation and the NV symmetry axis is 54.7$^\circ$. 
The red PL emitted from the NV centers in diamond is filtered by a long-pass filter (LPF-JGS1-12.7-650-300-625-675-1200) and collected by a PD. To enhance the sensitivity of the diamond magnetometer, a pair of magnetic flux concentrators made from high magnetic permeability materials(1J85) was adopted to improve the sensitivity of the MCG system. 
The geometrics of the flux concentrators was evaluated to balance the cardiac signal's attenuation with distance and the amplification factor of the flux concentrators. 

\section{Living rat specimen treatment}
The animal experiment was approved by  the Institutional Animal Care and Use Committee of the University of Science and Technology of China (USTCACUC26120223022).
In the animal experiment, the treatment process of the 313.5g living female rat is as followed. The rat underwent initial anesthesia by administering isoflurane. 
After the rat specimen fell into an unconscious state, an intraperitoneal injection injection of 20\% urethane with 2.5 mL dose for deeper anesthesia. 
10 minutes after the injection, the rat was adequately anesthetized and the installation process was performed. 
The rat was set on a multi-axis translation stage, which enabled a precise positioning from the NV diamond sensor in MCG measurement.
A custom-made acrylic fixing plate was designed and used for rat fixing and positioning. 

During the experiment, the diamond magnetometer was positioned directly above the rat's chest with about 5mm air gap, with the magnetic-sensitive axis of the diamond sensor aligned perpendicular to the plane of the rat chest. 
The duration of anesthesia was 12 hours after the intraperitoneal injection.
Three electrodes was fixed at the rat's body surface to record the ECG signal with the Lead II configuration, so that the MCG data and the ECG data could be compared in data analysis\cite{tontodonati2011improved}.
During the experiment, the heart rate of the rat was from 6.3 Hz to 7.9 Hz, corresponding to the speculated rat heart rate\cite{korecky1978normal}. 

\section{Optimization Simulation}
Simulation of the potential sensitivity enhancement is shown in the Figure \ref{SM_fig2}. 
The simulation was performed on the current system structure, where a diamond was clamped by two magnetic flux concentrators. 
In the assumptions required for the simulation of fluorescence collection efficiency, the excitation area in the diamond is considered as a uniform cylinder with a diameter of 40 $\mu m$. 
A series of uniformly spaced point light sources with a 360$^\circ$ divergence angle are used to simulate the behavior of color centers.
In the optimized simulation scheme, an additional mirror is placed perpendicular to the direction of laser incidence, and the sides of the magnetic flux concentrator are polished to an equivalent reflectivity of 0.8. 
For the simulation to evaluate the amplification factor of the magnetic flux concentration, the material's relative magnetic permeability is set to 40000. 
A uniform magnetic field is applied along the symmetry axis of the flux concentrators within a finite domain to conduct finite element static magnetism simulations, and the amplification factor of the magnetic flux concentration effect is calculated based on this background uniform magnetic field.

\clearpage
\bibliography{card.bib}

\end{document}